\documentclass[%
superscriptaddress,
twocolumn,
amsmath,amssymb,
aps,
prl,
]{revtex4-2}

\usepackage{graphicx}
\usepackage{dcolumn}
\usepackage{bm}
\usepackage{xcolor} 
\usepackage{hyperref}
\usepackage{amsmath}
\usepackage{amssymb}
\usepackage{mathtools}
\usepackage{comment}
\usepackage{float}
\usepackage[normalem]{ulem}

\newcommand{\omp}{\omega_{\rm p}}
\newcommand{\omo}{\omega_{\rm 0}}

\newcommand{\Gbx}{\Gamma\beta_x}
\newcommand{\Gby}{\Gamma\beta_y}
\newcommand{\mcN}{\mathcal{N}}
\newcommand{\gu}{\gamma_{\rm u}}

\newcommand{\me}{m_{\rm e}}

\newcommand{\eqb}{\begin{eqnarray}}
\newcommand{\eqe}{\end{eqnarray}}
\newcommand{\be}{\begin{eqnarray}}
\newcommand{\ee}{\end{eqnarray}}
\newcommand{\bi}{\begin{itemize}}
\newcommand{\ei}{\end{itemize}}

\newcommand{\llinear}{\ell_{\rm lin}}

\begin{document}


\title{Interaction of Strong Electromagnetic Waves with Unmagnetized Pair Plasmas}

\author{Navin Sridhar}
\email{navinsridhar@stanford.edu}
\affiliation{Department of Physics, Stanford University, 382 Via Pueblo Mall, Stanford, CA 94305, USA} 
\affiliation{Kavli Institute for Particle Astrophysics \& Cosmology, 452 Lomita Mall, Stanford, CA 94305, USA} 

\author{Emanuele Sobacchi}
\affiliation{Gran Sasso Science Institute, viale F. Crispi 7, L'Aquila, 67100, Italy}
\affiliation{INFN -- Laboratori Nazionali del Gran Sasso, via G. Acitelli 22, Assergi, 67100, Italy}

\author{Lorenzo Sironi}
\affiliation{Department of Astronomy and Columbia Astrophysics Laboratory, Columbia University, New York, NY 10027, USA}
\affiliation{Center for Computational Astrophysics, Flatiron Institute,  162 Fifth Avenue, New York, NY 10010, USA}

\author{Masanori Iwamoto}
\affiliation{Graduate School of System Informatics, Kobe University, 1-1 Rokkodai-cho, Nada-ku, Kobe 657-8501  Japan}
\affiliation{Yukawa Institute for Theoretical Physics, Kyoto University, Kitashirakawa-Oiwakecho, Sakyo-Ku, Kyoto 606-8502, Japan}

\author{Daniel Gro\v{s}elj}
\affiliation{Centre for mathematical Plasma Astrophysics, Department of Mathematics, KU Leuven, B-3001 Leuven, Belgium}

\author{Brandon K. Russell}
\affiliation{Department of Astrophysical Sciences, Princeton University, Princeton, New Jersey 08544, USA}




\date{\today}

\begin{abstract}
We investigate analytically and numerically the interaction of strong electromagnetic waves with unmagnetized pair plasmas. We show that the interaction is governed by a single nonlinearity parameter, $\varepsilon_{\rm p}$, defined as the ratio of the wave strength parameter to the wave frequency in units of the plasma frequency (with both frequencies measured in the plasma rest frame prior to the interaction). When $\varepsilon_{\rm p}<1$, the number of wavelengths that propagate through the plasma without attenuation from induced Compton scattering is approximately $\varepsilon_{\rm p}^{-2/3}$. This attenuation can imprint sub-structures as narrow as a few wavelengths on the pulse profile. When $\varepsilon_{\rm p}>1$, the electromagnetic pulse acts as a relativistic piston and drives a shock into the plasma. Our results establish a framework for the interaction of strong electromagnetic waves with pair plasmas, a process relevant for intense radio pulses from neutron stars and for next-generation pair plasma experiments at multi-petawatt laser facilities.

\end{abstract}

\maketitle


\textit{Introduction}---The interaction of strong electromagnetic (EM) waves with electron-proton plasmas has been extensively studied in the laboratory, where it is central to the development of plasma-based accelerators \cite{Esarey&Schroeder_09, Macchi+2013}. The discovery of fast radio bursts (FRBs)---bright extragalactic radio transients likely produced by magnetars \citep{Lorimer+07, Petroff+19, Cordes&Chatterjee19, Petroff+2022}---has renewed interest in the propagation of strong EM waves through astrophysical plasmas. Propagation effects could shape the spectral and temporal profiles of FRBs and determine whether FRBs can escape their sources \citep{Lyubarsky08, Nishiura+25a, Nishiura+25b, Nishiura+26, Beloborodov_22, Sobacchi+23, Iwamoto+23, Sobacchi+24a, Sobacchi+24b, Beloborodov_24, Sobacchi2025}. Plasmas in high-energy astrophysical environments---such as the magnetospheres of neutron stars and black holes---are loaded with electron-positron pairs \citep{goldreich_julian_69, Blandford&Znajek77}. This distinctive composition makes the propagation of strong EM waves different from that in laboratory electron-proton plasmas.

The interaction of EM waves with pair plasmas depends on a few parameters: (1)~The plasma frequency $\omp=\sqrt{8\pi n_0 e^2/\me}$, where $n_0$ is the proper density of the electrons ahead of the EM pulse. (2)~The wave frequency $\omo = 2\pi c/\lambda_0$, measured in the frame where the plasma ahead of the pulse is at rest. (3)~The Lorentz invariant wave strength parameter $a_0 = eE_0/\me c \omo$, where $E_0$ is the amplitude of the wave electric field. When the EM wave is strong ($a_0>1$), particles oscillate at relativistic velocities when illuminated by the wave \citep{Gunn&Ostriker_71, LandauLifshitz1975}.

A quantitative understanding of the interaction between strong EM waves and pair plasmas has not been achieved. Ref.~\cite{Sobacchi+24b} showed analytically that the current carried by the plasma is a linear function of the wave electric field when $a_0\ll\omo/\omp$ (instead, in electron-proton plasmas, the current is linear when $a_0\ll 1$ \citep{Kruer+19}). Then, this suggests that in pair plasmas, nonlinear propagation effects could be governed by the parameter $\varepsilon_{\rm p}\equiv a_0\omp/\omo$. Kinetic simulations have shown that underdense pair plasmas can fully reflect EM waves when $\varepsilon_{\rm p}\gtrsim1$ \citep[e.g., Ref.~][]{Tangtartharakul+25}. However, it is unclear whether EM waves can propagate indefinitely through the plasma for all values of $\varepsilon_{\rm p}<1$, or whether they could eventually be absorbed. Previous works have shown that the propagation of EM waves in pair plasmas can be severely affected by induced Compton scattering  \citep{Lyubarsky08, Edwards+16, Nishiura+25a, Nishiura+25b, Nishiura+26, Lyubarsky2019b, Ghosh+22, LyutikovGurarie2025, Kamijima+26}. However, except for the analytical calculation of the scattering rate presented in Ref.~\citep{Lyubarsky2019b, Iwamoto&Ioka_26}, these works were limited to weak EM waves ($a_0\ll 1$).

In this \textit{Letter}, we combine analytical theory and kinetic simulations to study the propagation of strong ($a_0>1$), monochromatic EM pulses in collisionless cold pair plasmas across the $\varepsilon_{\rm p}<1$ and $\varepsilon_{\rm p}>1$ regimes. We analytically derive the growth rate of the induced Compton scattering of strong waves when $\varepsilon_{\rm p}<1$. Our large computational domain allows us to show that the number of wavelengths that propagate through the plasma without attenuation is approximately $\varepsilon_{\rm p}^{-2/3}$, in agreement with the analytical growth rate of induced scattering. We also show that the EM pulse acts as a relativistic piston and drives a shock into the plasma when $\varepsilon_{\rm p}>1$.

\textit{Analytical theory}---We consider a linearly polarized, monochromatic EM pulse that propagates along $+\hat{x}$ in a cold pair plasma. In the frame where the plasma ahead of the pulse is at rest, the wave frequency is $\omega_0$, and the wave vector is $k_0$. The electric field of the wave is directed along $\hat{y}$ and the magnetic field along $\hat{z}$. We assume that all physical quantities depend only on $x$ and $t$. We adopt a two-fluid model for the plasma. Electrons and positrons move in the $xy$ plane. They have the same number densities, the same velocities along $\hat{x}$, and opposite velocities along $\hat{y}$. The equations governing the evolution of the system can be presented as \cite{Gruzinov2019, Sobacchi+24a}
\begin{align}
\label{eq:main1}
& \frac{\partial}{\partial t}\left(\gamma n\right) + \frac{\partial}{\partial x}\left(nu_x\right) = 0 \\
\label{eq:main2}
& \frac{\partial u_x}{\partial t} + \frac{\partial \gamma}{\partial x} = 0 \\
\label{eq:main3}
& \frac{\partial^2 a}{\partial t^2} - \frac{\partial^2 a}{\partial x^2} +\omp^2\frac{n}{n_0}a = 0 \;,
\end{align}
where $u_x$ and $\gamma=\sqrt{1+a^2+u_x^2}$ are the $x$ component of the four-velocity, and the Lorentz factor of electrons, and $n$ is their proper density. The $y$ component of the four-velocity is $u_y=a$, and the vector potential of the EM wave is given by $\me a/e$. Throughout this section, we work in units where the speed of light is $c=1$.

The dispersion relation of the wave in the plasma is $\omo^2 = \omp^2+k_0^2$. Since the wave is superluminal ($\omo> k_0$), it is convenient to work in the reference frame that moves with velocity $k_0/\omega_0$ along $+\hat{x}$. In this frame (termed the `wave frame' hereafter), the wave frequency is $\omega=\sqrt{\omega_0^2-k_0^2}$ and the wave vector vanishes \citep[][]{Clemmow_1974}. Then, the unperturbed solution of Eqs.~\eqref{eq:main1}-\eqref{eq:main3} is independent of $x$. From Eqs.~\eqref{eq:main1} and \eqref{eq:main2} we find, respectively, $\gamma n=\gamma_0n_0$ and $u_x=-u_0$, where $u_0=k_0/\omega$ and $\gamma_0=\omo/\omega$. From Eq.~\eqref{eq:main3} we find $\partial^2 a/\partial t^2+ \omp^2 a/ \sqrt{1+a^2/\gamma_0^2}=0$, whose solution is denoted as $\bar{a}(t)$. When $a_0\ll\gamma_0$, we can approximate $\bar{a}=(a_0/2)[\exp(-{\rm i}\omp t)+ \exp({\rm i}\omp t)]$, which implies $\omega=\omp$ and $\gamma_0=\omo/\omp$.

To study the stability of the unperturbed solution, we define $\gamma n=\gamma_0n_0+\delta\rho$, $u_x = -u_0+\delta u$, and $a=\bar{a}+\delta a$ (where $\delta \rho$, $\delta u$, $\delta a$ are small perturbations). Substituting these expressions into Eqs.~\eqref{eq:main1}-\eqref{eq:main3}, and neglecting the quadratic terms in the perturbations, we find
\begin{align}
\label{eq:pert1}
& \frac{\rm D}{{\rm D}t} \left(\frac{\delta\rho}{\gamma_0n_0}\right) + \frac{1+\bar{a}^2}{\gamma_0^3} \frac{\partial\delta u}{\partial x} + \frac{\bar{a}}{\gamma_0^2} \frac{\partial\delta a}{\partial x} = 0 \\
\label{eq:pert2}
& \frac{\rm D}{{\rm D}t} \left(\frac{\delta u}{\gamma_0}\right) + \frac{\bar{a}}{\gamma_0^2}\frac{\partial\delta a}{\partial x} = 0 \\
\label{eq:pert3}
& \frac{\partial^2\delta a}{\partial t^2} - \frac{\partial^2\delta a}{\partial x^2} + \omp^2\delta a + \omp^2 \bar{a} \left( \frac{\delta\rho}{\gamma_0 n_0} + \frac{\delta u}{\gamma_0} \right) = 0 \;,
\end{align}
where we assumed $a_0\ll\gamma_0$ and defined
\begin{equation}
\frac{\rm D}{{\rm D}t}\equiv \frac{\partial}{\partial t} - \left(1-\frac{1+\bar{a}^2}{2\gamma_0^2}\right) \frac{\partial}{\partial x} \;.
\end{equation}

The nonlinear terms of Eqs.~\eqref{eq:pert1}-\eqref{eq:pert3} couple to different modes. We look for a solution of the form $\delta\rho/\gamma_0 n_0 = Q \exp[{\rm i}Kx - {\rm i}\Omega t]$, $\delta u/\gamma_0 = U \exp[{\rm i}Kx - {\rm i}\Omega t]$, $\delta a/a_0 = A_+ \exp[{\rm i}Kx - {\rm i}\left(\Omega +\omega_{\rm P}\right) t] + A_- \exp[{\rm i}Kx - {\rm i}\left(\Omega -\omega_{\rm P}\right) t]$. Substituting these expressions into Eqs.~\eqref{eq:pert1}-\eqref{eq:pert3}, we find
\begin{align}
\label{eq:four1}
D_{\rm v} Q & = \frac{1+a_0^2/2}{\gamma_0^2} KU + \frac{a_0^2}{2\gamma_0^2} K \left(A_+ + A_-\right) \\
\label{eq:four2}
D_{\rm v} U & = \frac{a_0^2}{2\gamma_0^2} K \left(A_+ + A_-\right) \\
\label{eq:four3}
D_\pm A_\pm & = \frac{\omp^2}{2} \left(Q+U \right) \;,
\end{align}
where we defined
\begin{align}
D_{\rm v} & \equiv \Omega + \left(1- \frac{1+a_0^2/2}{2\gamma_0^2} \right) K \\
D_\pm & \equiv \left(\Omega\pm \omp \right)^2 -K^2-\omp^2 \;.
\end{align}
The dispersion relation can be determined from Eqs.~\eqref{eq:four1}-\eqref{eq:four3}. It can be presented as
\begin{equation}
\label{eq:DR}
D_{\rm v}^2 = \frac{\omp^2a_0^2}{2\gamma_0^2} K\left(\Omega+K\right) \left(\frac{1}{D_+}+\frac{1}{D_-}\right) \;.
\end{equation}

The growth rate of the instability, $\Delta\Omega_\pm$, can be determined by substituting $K=K_\pm$ and $\Omega=\Omega_\pm+\Delta\Omega_\pm$ into Eq.~\eqref{eq:DR}. Here, $K_\pm$ and $\Omega_\pm$ are determined by the conditions $D_{\rm v}=0$ and $D_\pm=0$. The most unstable wave number is given by
\begin{equation}
\label{eq:Kstar}
K_\pm = \mp \frac{2\gamma_0^2\omega_{\rm P}}{1+a_0^2/2} \;,
\end{equation}
and the growth rate of the instability is given by
\begin{equation}
\label{eq:Ostar}
\left(\Delta\Omega_\pm\right)^3 = \pm \frac{a_0^2}{4\gamma_0^2}\omp^3 \;.
\end{equation}
Eqs.~\eqref{eq:Kstar}-\eqref{eq:Ostar} are valid for $a_0>1$ inasmuch as $a_0\ll\gamma_0$---or, equivalently, $\varepsilon_{\rm p}\equiv a_0\omp/\omo\ll 1$. Notably, this also recovers the induced scattering rate of weak waves ($a_0\ll 1$) propagating in cold pair plasmas calculated by Refs.~\citep{Nishiura+25b, LyutikovGurarie2025}. We cannot compare Eqs.~\eqref{eq:Kstar}-\eqref{eq:Ostar} with the scattering rate of strong waves calculated by Ref.~\citep{Lyubarsky2019b} because they considered waves with a broadband spectrum.

\begin{figure*}
\includegraphics[width=0.34\textwidth]{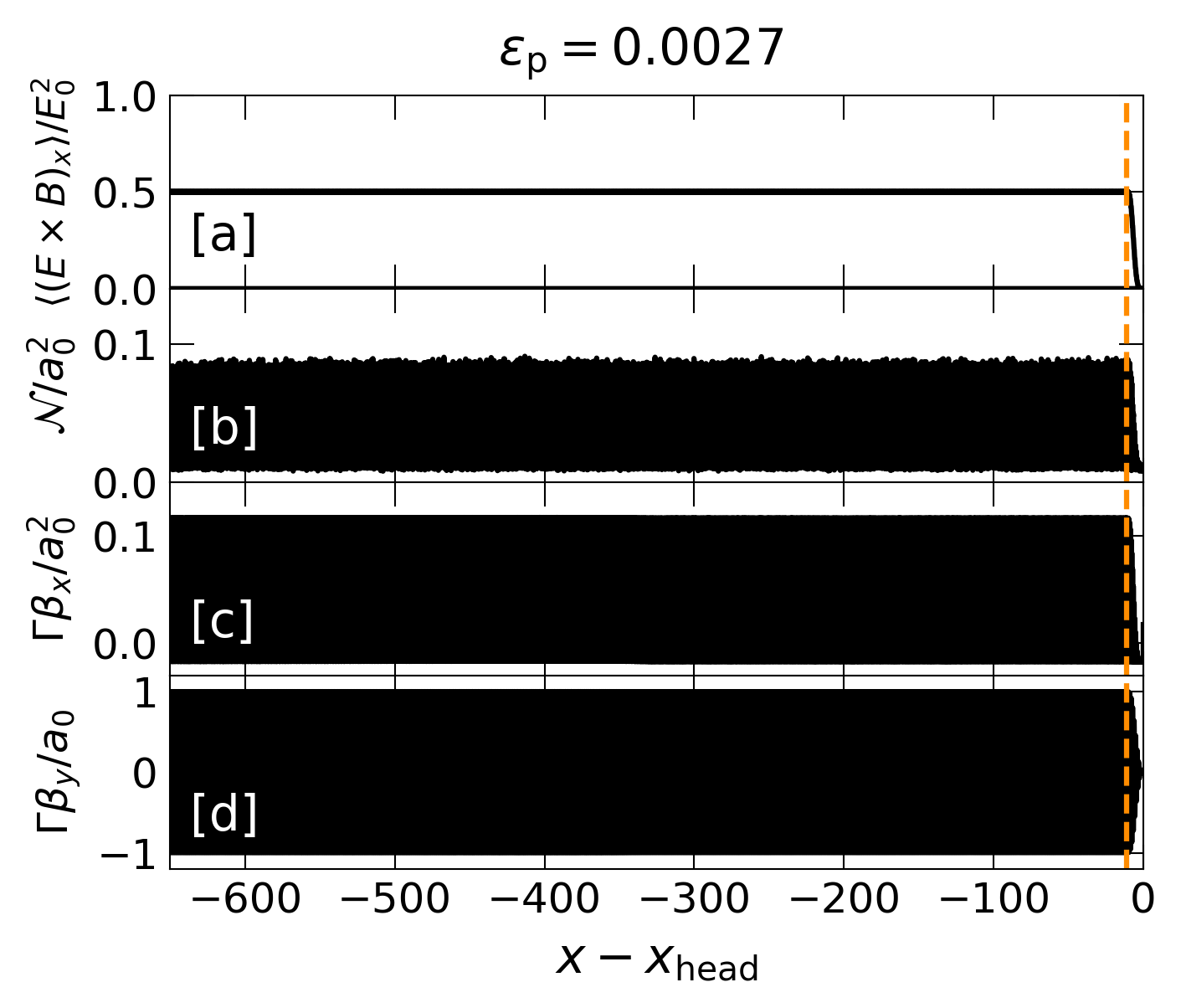}
\includegraphics[trim={0.8cm 0 0 0},clip, width=0.315\textwidth]{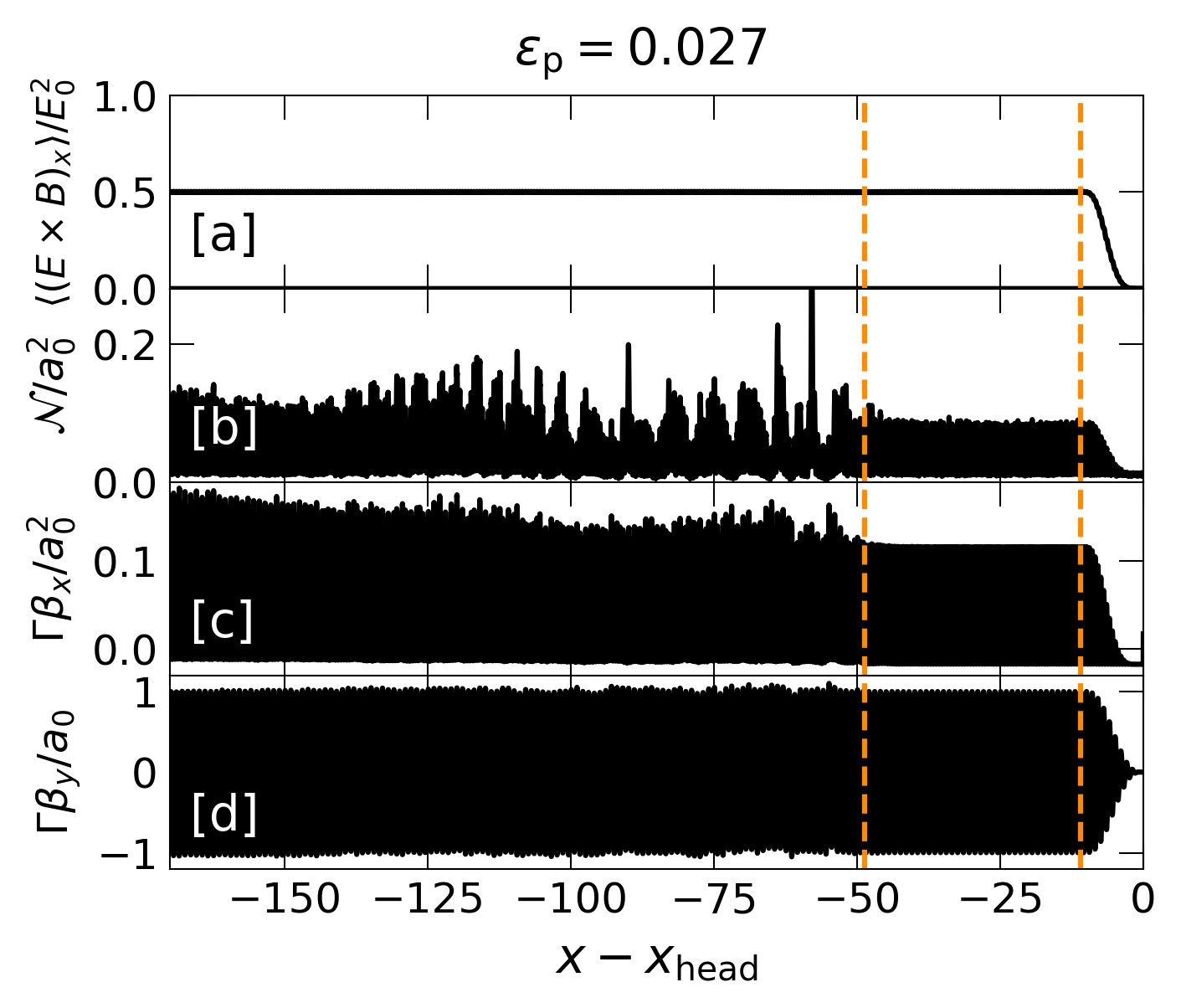}
\includegraphics[trim={1.06cm 0 0 0},clip, width=0.315\textwidth]{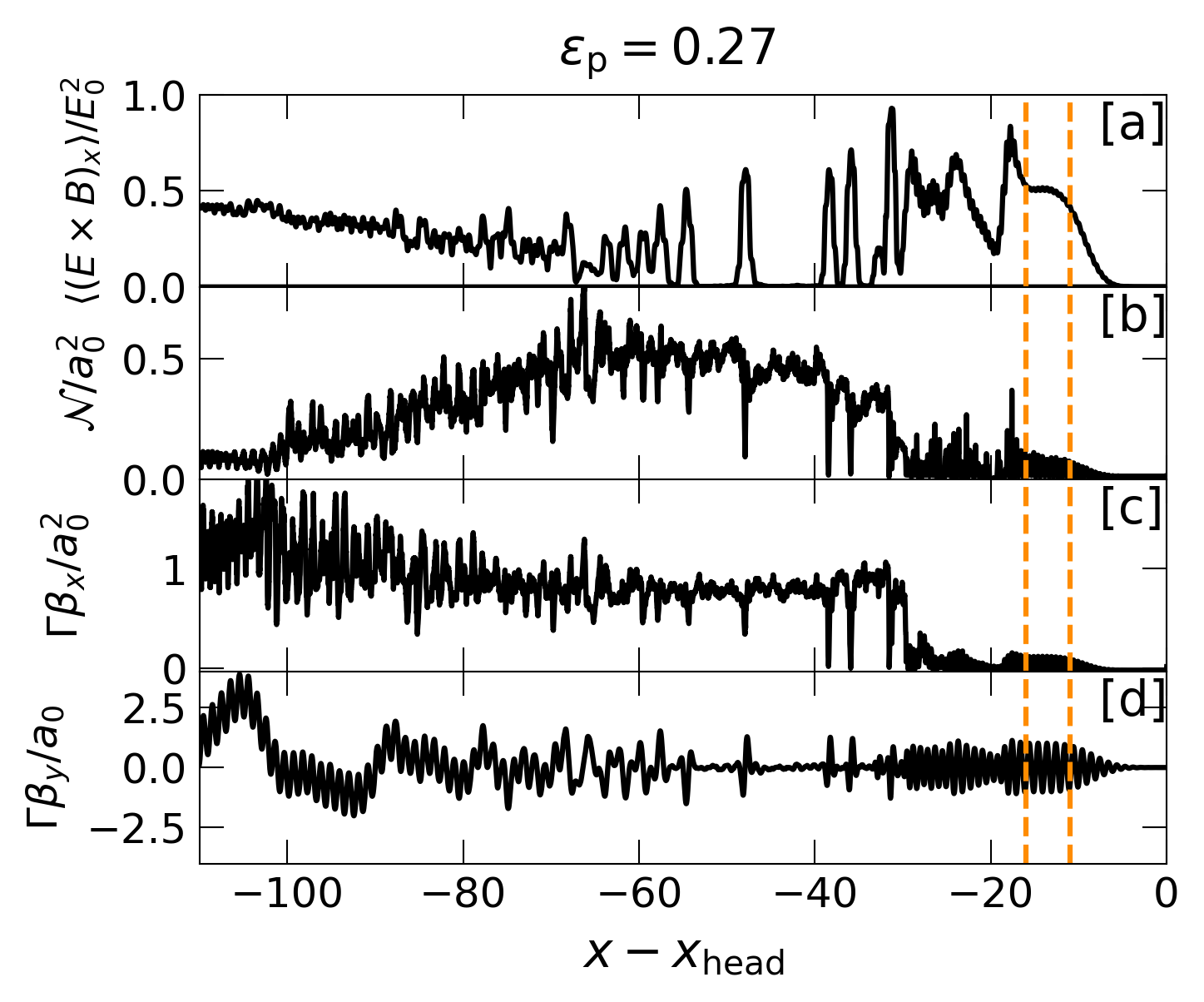}
\caption{\label{fig:spatial_map_epsilon} EM pulse-plasma interaction region at time $t\omega_0=45715$. The simulations have the same wave strength parameter $a_0=10$, but different nonlinearity parameters $\varepsilon_{\rm p} = 0.0027$ ($\omo/\omp=3700$; left column), $\varepsilon_{\rm p} =0.027$ ($\omo/\omp=370$; middle column), $\varepsilon_{\rm p} = 0.27$ ($\omo/\omp=37$; right column). [a] Phase-averaged Poynting flux along the $\hat{x}$ direction  (normalized by the square of the vacuum electric field in the simulation frame), [b] particle number density in units of the upstream density (normalized by $a_0^2$), and cell-averaged four-velocity of the particles [c] along the $\hat{x}$ direction (normalized by $a_0^2$) and [d] along the $\hat{y}$ direction (normalized by $a_0$). The distance from the head of the EM pulse, $x-x_{\rm head}$, is expressed in units of the vacuum wavelength in the simulation frame. The orange dashed vertical lines enclose the linear propagation length, $\llinear$.}
\end{figure*}

In the wave frame, the distance from the head of the EM pulse where induced scattering significantly affects the pulse profile is approximately $1/|\Delta\Omega_\pm|\simeq \varepsilon_{\rm p}^{-2/3}\omp^{-1}$. In the reference frame where the plasma ahead of the pulse is at rest, the distance is
\begin{equation}
\label{eq:mainresult}
\llinear \simeq \frac{1}{\gamma_0|\Delta\Omega_\pm|} \simeq \varepsilon_{\rm p}^{-2/3} \lambda_0 \;.
\end{equation}
The Lorentz invariant ratio $\llinear/\lambda_0$ is governed by the nonlinearity parameter $\varepsilon_{\rm p}$. Hereafter, $\llinear$ is termed `linear propagation length'. Beyond the linear propagation length, the EM pulse is attenuated because the energy of the forward-propagating pump wave is transferred into the upper and lower back-scattered sidebands and into plasma density/velocity fluctuations, as described by the dispersion relation in Eq.~\eqref{eq:DR}.

\textit{Numerical simulations}---Our numerical simulations are performed with the particle-in-cell code {\tt OSIRIS} \cite{Fonseca+02}. We initialize a computational domain that contains a linearly polarized, monochromatic EM pulse propagating along $+\hat{x}$ in vacuum. The electric and magnetic fields of the wave are directed along $\hat{y}$ and $\hat{z}$, respectively. The computational domain along $\hat{x}$ is chosen to encompass the entire pulse, whose duration is adapted to capture the nonlinear propagation effects. The amplitude of the EM pulse is gradually increased over 10 wavelengths to its maximum value (corresponding to a strength parameter $a_0$). The EM pulse eventually encounters the `upstream' plasma that drifts towards $-\hat{x}$ with a Lorentz factor $\gu$ in the simulation frame. We adopt $\gu=1$ or $2$ and test the frame-independence of our results (see End Matter). The number density increases gradually as the plasma enters the computational window at the right boundary. After passing through the pulse, the particles exit the computational domain on the left through an open/absorbing boundary modeled as a perfectly matched layer \cite{Vay_2002}. The computational domain drifts along $+\hat{x}$ at the speed of light to follow the pulse's propagation. Along $\hat{y}$, the domain contains 5 cells (with periodic boundaries) to accommodate the shape of the macroparticle and the wide stencil of our EM field solver (more in the End Matter). The particles are initialized as a Maxwellian distribution with a normalized temperature in the range $\theta_0 = k_{\rm B}T/\me c^2 = 10^{-6} - 10^{-2}$, with $\theta_0 = 10^{-4}$ being our fiducial case (see the End Matter). We explored a wide range of nonlinearity parameters ($\varepsilon_{\rm p}\sim 10^{-4}-30$) by varying $a_0 = 0.01-300$ and $\omo/\omp = 3.7-3700$ (measured in the upstream plasma rest frame).

\begin{figure*}
\includegraphics[width=0.9\textwidth]{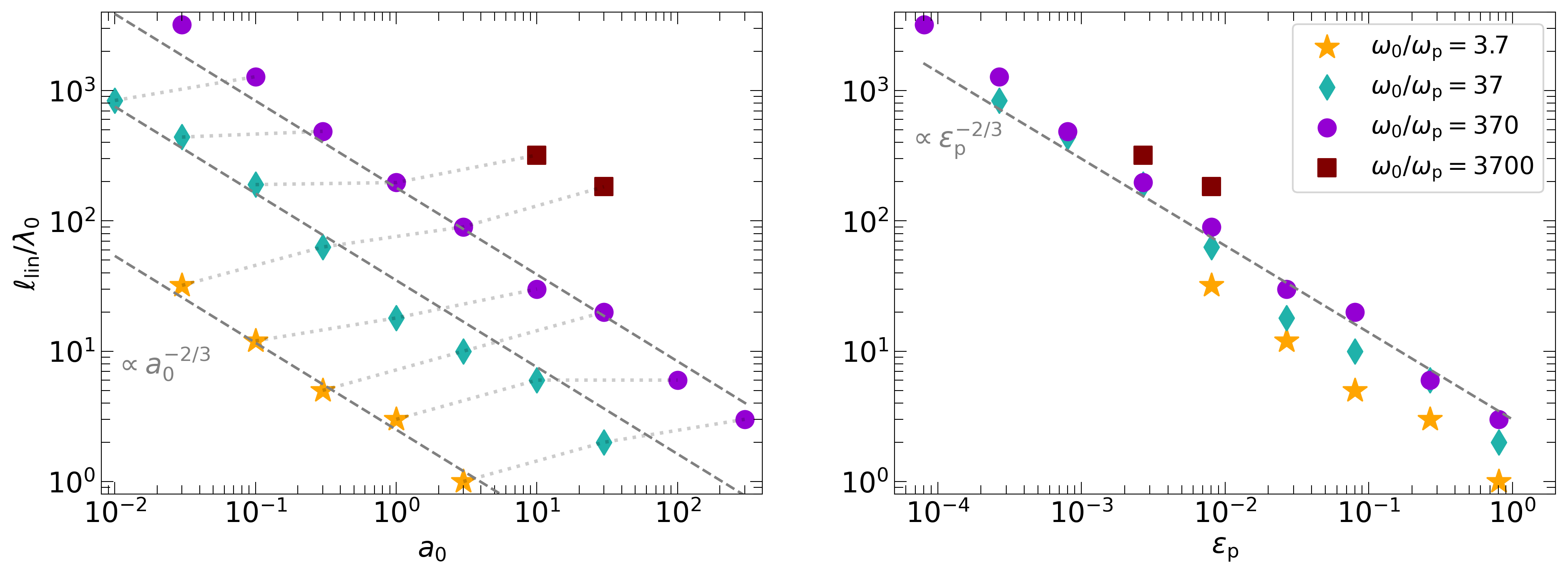}
\caption{\label{fig:lin_prop_length}
Linear propagation length in units of the vacuum wavelength as a function of $a_0$ (left panel) and $\varepsilon_{\rm p}$ (right panel). Different markers denote simulations with different $\omo/\omp$ (see legend). The dashed lines in the left and right panels represent the $a_0^{-2/3}$ and $\varepsilon_{\rm p}^{-2/3}$ scalings, respectively. The dotted curves in the left panel connect markers with the same $\varepsilon_{\rm p}$.}
\end{figure*}

\textit{Weakly nonlinear regime ($\varepsilon_{\rm p}<1$)}---In Fig.~\ref{fig:spatial_map_epsilon}, we show the spatial structure of the EM pulse-plasma interaction region in the simulation frame ($\gu=2$) at time $t\omo=45715$. We adopt $a_0=10$ and vary the nonlinearity parameter as follows: $\varepsilon_{\rm p}=0.0027$ ($\omo/\omp=3700$; left column), $\varepsilon_{\rm p}=0.027$ ($\omo/\omp=370$; middle column) and $\varepsilon_{\rm p}=0.27$ ($\omo/\omp=37$; right column). In each column, we show [a] the phase-averaged Poynting flux along the $\hat{x}$ direction, [b] the particle number density in units of the upstream density ($\mcN$), and the cell-averaged four-velocity of the particles [c] along the $\hat{x}$ direction ($\Gbx$) and [d] along the $\hat{y}$ direction ($\Gby$).

Nonlinear propagation effects produce large-amplitude fluctuations of all physical quantities. When $\varepsilon_{\rm p}$ increases, the fluctuations become larger and appear closer to the head of the EM pulse. When $\varepsilon_{\rm p}=0.0027$, the onset of nonlinearity is barely discernible after the pulse propagates $\sim600$ wavelengths. When $\varepsilon_{\rm p}=0.027$, fluctuations of $\mcN$, $\Gbx$, and $\Gby$ are easily noticeable for $x-x_{\rm head}<-50$, but the phase-averaged Poynting flux is weakly affected by nonlinearity at this time. When $\varepsilon_{\rm p}=0.27$, the EM pulse breaks into sub-structures as short as a few wavelengths that appear for $x-x_{\rm head}<-15$.

The orange dashed vertical lines in Fig.~\ref{fig:spatial_map_epsilon} enclose the `linear propagation length' before the onset of nonlinear effects. We estimate it as $\llinear=x_1-x_2$, where $x_1$ is the end of the ramp-up phase of the EM pulse, and $x_2$ is the position where $\Gby$ departs from the motion of a test particle in a vacuum EM wave. More precisely, $x_2$ is defined as the first instance in which $|\Gby-(\Gby)_{\rm vac}|>\varepsilon_{\rm thr}a_0$. Here we introduce $(\Gby)_{\rm vac}\equiv eE_y/\me c \gu(1-\beta_{\rm u})\omo$, where $E_y$ is measured in the simulation frame and its phase is shifted by $\pi/2$, and $\gu(1-\beta_{\rm u})\omo$ is the vacuum frequency in the simulation frame. We adopted a threshold $\varepsilon_{\rm thr}=10^{-6}$ and verified that $\llinear$ is nearly independent of the threshold for $\varepsilon_{\rm thr}<10^{-6}$.

In Fig.~\ref{fig:lin_prop_length}, we demonstrate that the linear propagation length is $\llinear/\lambda_0\simeq\varepsilon_{\rm p}^{-2/3}$, as predicted by Eq.~\eqref{eq:mainresult}. In the left panel, we show that $\llinear/\lambda_0$ is proportional to $a_0^{-2/3}$ when $\omo/\omp$ is fixed (see dashed lines), and increases for larger $\omo/\omp$ when $a_0$ is fixed. In the right panel, we show that the expression $\llinear/ \lambda_0 \simeq \varepsilon_{\rm p}^{-2/3}$ captures the results of the simulations for all combinations of $a_0$ and $\omo/\omp$ (see dashed line). We do not show $\llinear/\lambda_0$ for simulations with $\varepsilon_{\rm p}>1$ because, as we discuss below, not a single wavelength of the initial EM pulse propagates into the plasma.

\begin{figure}[t]
\includegraphics[width=0.45\textwidth]{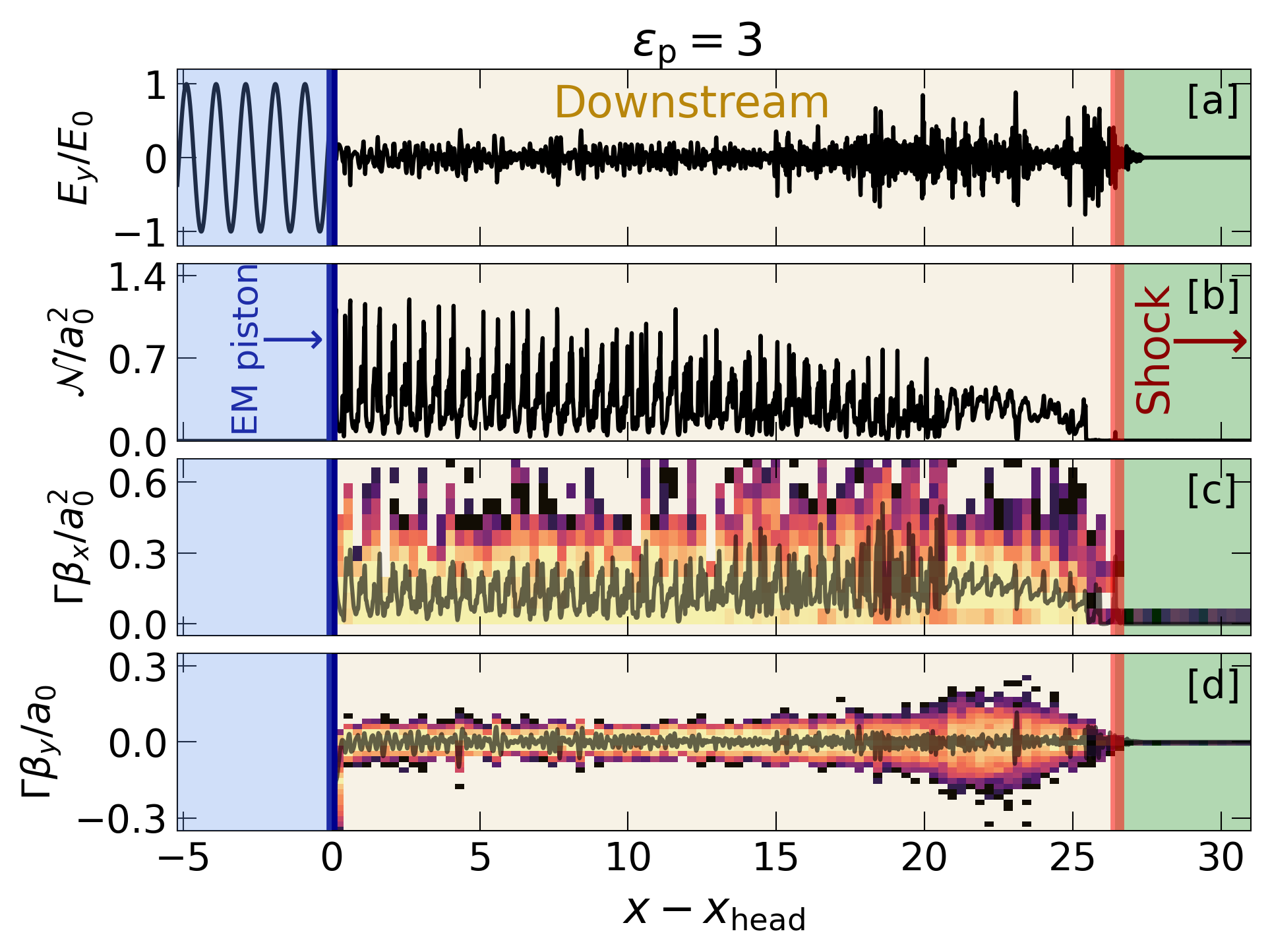}
\caption{\label{fig:spatial_map_epsilon3}
EM pulse-plasma interaction region at time $t\omega_0=13929$. We adopt a nonlinearity parameter $\varepsilon_{\rm p}=3$ ($a_0=30$, $\omo/\omp=10$). [a] Oscillating electric field (normalized by the vacuum electric field), [b] particle number density in units of the upstream density (normalized by $a_0^2$), and cell-averaged four-velocity of the particles [c] along the $\hat{x}$ direction (normalized by $a_0^2$) and [d] along the $\hat{y}$ direction (normalized by $a_0$). In panels [c-d], we show the distribution in momentum space (where bright colors denote higher density) and the cell-averaged $\Gbx$ and $\Gby$ (gray curves). The shaded regions denote the vacuum EM pulse (blue), the shock downstream (yellow), and the upstream (green).
}
\end{figure}

\textit{Highly nonlinear regime ($\varepsilon_{\rm p}> 1$)}---In Fig.~\ref{fig:spatial_map_epsilon3}, we show the spatial structure of the EM pulse-plasma interaction region in the simulation frame ($\gu=1$) at time $t\omo=13929$. The nonlinearity parameter is $\varepsilon_{\rm p}=3$ ($a_0=30$, $\omo/\omp=10$). The EM wave does not propagate through the plasma (panel [a]; see also Ref.~\citep{Tangtartharakul+25}). Instead, the radiation pressure acts as a piston (blue vertical line) and drives a shock ahead (red vertical line). The lack of charge separation in pair plasmas makes the shock structure distinct from the electrostatic `double-layer' shocks formed during the interaction of high-intensity lasers with electron-proton plasmas \cite{Palmer+11, Haberberger+12}.

In the piston frame, the wave radiation pressure balances the momentum flux of the incoming plasma. Therefore, the Lorentz factor of the piston in the upstream frame, $\gamma_{\rm P}=(1-\beta_{\rm P}^2)^{-1/2}$, is given by the implicit equation $\sigma_{\rm emw}(1-\beta_{\rm P}) = 2\gamma_{\rm P}^2\beta_{\rm P}^2 (1+\beta_{\rm P})$ \citep{Robinson+09, Schlegel+09}; here, $\sigma_{\rm emw}\equiv a^2(t)\omo^2/\omp^2$ is the instantaneous ratio of the EM wave energy density to plasma rest mass energy density. When $\sigma_{\rm emw}\gg 1$, the Lorentz factor of the piston is $\gamma_{\rm P}\approx (\sigma_{\rm emw}/8)^{1/4}$. In the piston frame, the number density does not change when the particles are reflected by the piston, and the four-velocity along the $\hat{x}$ direction is reversed. In the upstream frame, the number density and the four-velocity of the reflected particles are given by $\mcN \simeq \Gbx \approx \sqrt{\sigma_{\rm emw}/2}$ \citep{Robinson+09, Schlegel+09}.

These analytical estimates are broadly consistent with the numerical results presented in Fig.~\ref{fig:spatial_map_epsilon3} (panels [b-c]) and with additional simulations where we vary $1<\varepsilon_{\rm p}<30$. In the downstream region, both $\mcN$ and $\Gbx$ exhibit compressive oscillations with frequency $\sim2\omega_0$ because $\sigma_{\rm emw}$ is proportional to $a^2(t)=a_0^2\cos^2{(\omo t)}$. The shock thermalizes the upstream plasma, as shown by the downstream particle distribution in momentum space (panel [c]). The depletion of the wave electric field in the downstream region implies $\Gby\ll a_0$ (panel [d]). Overall, in the $\varepsilon_{\rm p}\gtrsim1$ regime, the EM wave is irreversibly absorbed by the plasma; the attenuation here proceeds via mechanical work done by the EM wave piston on the plasma.

\textit{Conclusions}---In this work, we have studied the interaction of linearly polarized strong EM waves ($a_0\gg1$) with unmagnetized pair plasmas. This interaction is governed by the nonlinearity parameter $\varepsilon_{\rm p}\equiv a_0\omp/\omo$ \citep{Sobacchi+24b}, and is expected to be different from the interaction with electron-proton plasmas, where nonlinear effects are governed by $a_0$ \citep{Kruer+19}. We have identified two regimes of EM wave-pair plasma interaction.

(1) \textit{Weakly nonlinear regime ($\varepsilon_{\rm p}<1$)}. The number of wavelengths that propagate without disturbance through the plasma is limited to $\varepsilon_{\rm p}^{-2/3}$. The propagation of longer EM pulses is inhibited by induced Compton scattering, whose growth rate and most unstable wavenumber are analytically derived and numerically verified in the regime $a_0\gg1$.

(2) \textit{Highly nonlinear regime ($\varepsilon_{\rm p}> 1$)}. The wave does not propagate through the plasma, and its leading edge acts as a relativistic piston that drives a shock ahead. The shock-thermalized downstream particles exhibit large-amplitude oscillations at frequency $\sim 2\omo$ in their density and longitudinal four-velocity, whereas their transverse four-velocity nearly vanishes because the wave electric field is attenuated.

Our results have implications for next-generation pair plasma experiments at multi-petawatt lasers and high-energy accelerator facilities \citep{Chen&Fiuza_23, Arrowsmith+24}, and for the propagation of highly coherent astrophysical pulses---such as FRBs and giant radio pulses---in the outer magnetospheres of neutron stars.
At large distances from a magnetar ($\gtrsim10^{12}$\,cm), the local cyclotron frequency is expected to be much smaller than the $\sim$GHz EM wave frequency. If the wind magnetization parameter (i.e., the ratio of the electromagnetic to kinetic luminosities) drops below unity at large distances due to field dissipation \citep{Metzger+19, Beloborodov_20a, Cerutti_20}, our unmagnetized treatment of EM wave-plasma interaction can be applicable.
Our framework can provide the minimum radius within the aforementioned regime, beyond which an FRB-like pulse with a given luminosity and duration must be emitted to escape the system. Further, we show that propagation effects due to induced Compton scattering can imprint narrow features on the pulse---as short as a few wavelengths, thus offering a physically grounded alternative to the traditional causality-based emission-size arguments. The multi-dimensional effects, the impact of a strong background magnetic field, and the broadband nature of the EM pulse will be investigated in future work.

\textit{Acknowledgments---}This work benefits from discussions with Roger Blandford, Damiano Fiorillo, Fei Li, Anatoly Spitkovsky, Frank Tsung, and Xinlu Xu. N.S. acknowledges support from the Simons Foundation (grant MP-SCMPS-00001470). E.S. acknowledges support from a Rita Levi Montalcini fellowship. L.S. acknowledges support from the DoE Early Career Award DE-SC0023015, NASA ATP award \#80NSSC24K1826, a grant from the Simons Foundation (MP-SCMPS-00001470), the Multimessenger Plasma Physics Center (MPPC) grant PHY-2206609. D.G.~is supported by the Research Foundation--Flanders (FWO) Senior Postdoctoral Fellowship 12B1424N. B.K.R acknowledges support from NSF grant No. 2512021. This research was supported in part by the NSF grant PHY-2309135 to the Kavli Institute for Theoretical Physics (KITP). Some of the computing for this project was performed on the \textit{Sherlock} cluster at Stanford University. We thank Stanford University and the Stanford Research Computing Center for providing computational resources and support that contributed to these research results.

\textit{Data availability---}The data that support the findings of this article are not publicly available upon publication because it is not technically feasible and/or the cost of preparing, depositing, and hosting the data would be prohibitive within the terms of this research project. The data are available from the authors upon reasonable request.


\bibliography{aps}

@ARTICLE{Sobacchi2025,
       author = {{Sobacchi}, Emanuele},
        title = "{Absorption of strong electromagnetic waves in magnetized pair plasmas}",
      journal = {\pre},
         year = 2025,
        month = dec,
       volume = {112},
       number = {6},
          eid = {065208},
        pages = {065208},
          doi = {10.1103/1b36-qn66},
archivePrefix = {arXiv},
       eprint = {2512.05281},
 primaryClass = {astro-ph.HE},
       adsurl = {https://ui.adsabs.harvard.edu/abs/2025PhRvE.112f5208S}
}

@ARTICLE{Lyubarsky2019b,
       author = {{Lyubarsky}, Yuri},
        title = "{Interaction of the electromagnetic precursor from a relativistic shock with the upstream flow - II. Induced scattering of strong electromagnetic waves}",
      journal = {\mnras},
         year = 2019,
        month = nov,
       volume = {490},
       number = {1},
        pages = {1474-1478},
          doi = {10.1093/mnras/stz2712},
       adsurl = {https://ui.adsabs.harvard.edu/abs/2019MNRAS.490.1474L}
}

@ARTICLE{Petroff+2022,
       author = {{Petroff}, E. and {Hessels}, J.~W.~T. and {Lorimer}, D.~R.},
        title = "{Fast radio bursts at the dawn of the 2020s}",
      journal = {\aapr},
         year = 2022,
        month = dec,
       volume = {30},
       number = {1},
          eid = {2},
        pages = {2},
          doi = {10.1007/s00159-022-00139-w},
archivePrefix = {arXiv},
       eprint = {2107.10113},
 primaryClass = {astro-ph.HE},
       adsurl = {https://ui.adsabs.harvard.edu/abs/2022A&ARv..30....2P}
}

@article{Beloborodov_22,
	author = {Beloborodov, Andrei M.},
	doi = {10.1103/physrevlett.128.255003},
	issn = {1079-7114},
	journal = {Physical Review Letters},
	month = jun,
	number = {25},
	publisher = {American Physical Society (APS)},
	title = {Scattering of Ultrastrong Electromagnetic Waves by Magnetized Particles},
	url = {http://dx.doi.org/10.1103/PhysRevLett.128.255003},
	volume = {128},
	year = {2022}}

@article{Beloborodov_24,
	adsurl = {https://ui.adsabs.harvard.edu/abs/2024ApJ...975..223B},
	archiveprefix = {arXiv},
	author = {{Beloborodov}, Andrei M.},
	doi = {10.3847/1538-4357/ad698c},
	eid = {223},
	eprint = {2307.12182},
	journal = {\apj},
	month = nov,
	number = {2},
	pages = {223},
	primaryclass = {astro-ph.HE},
	title = {{Damping of Strong GHz Waves near Magnetars and the Origin of Fast Radio Bursts}},
	volume = {975},
	year = 2024}

@article{Sobacchi+24a,
	adsurl = {https://ui.adsabs.harvard.edu/abs/2024A&A...690A.332S},
	archiveprefix = {arXiv},
	author = {{Sobacchi}, E. and {Iwamoto}, M. and {Sironi}, L. and {Piran}, T.},
	doi = {10.1051/0004-6361/202451725},
	eid = {A332},
	eprint = {2409.10732},
	journal = {\aap},
	month = oct,
	pages = {A332},
	primaryclass = {astro-ph.HE},
	title = {{Escape of fast radio bursts from magnetars}},
	volume = {690},
	year = 2024}

@article{Sobacchi+24b,
	adsurl = {https://ui.adsabs.harvard.edu/abs/2024PhRvR...6d3213S},
	archiveprefix = {arXiv},
	author = {{Sobacchi}, Emanuele and {Iwamoto}, Masanori and {Sironi}, Lorenzo and {Piran}, Tsvi},
	doi = {10.1103/PhysRevResearch.6.043213},
	eid = {043213},
	eprint = {2409.04127},
	journal = {Physical Review Research},
	month = nov,
	number = {4},
	pages = {043213},
	primaryclass = {astro-ph.HE},
	title = {{Propagation of strong electromagnetic waves in tenuous plasmas}},
	volume = {6},
	year = 2024}

@article{Sobacchi+23,
	adsurl = {https://ui.adsabs.harvard.edu/abs/2023ApJ...943L..21S},
	archiveprefix = {arXiv},
	author = {{Sobacchi}, Emanuele and {Lyubarsky}, Yuri and {Beloborodov}, Andrei M. and {Sironi}, Lorenzo and {Iwamoto}, Masanori},
	doi = {10.3847/2041-8213/acb260},
	eid = {L21},
	eprint = {2210.08754},
	journal = {\apjl},
	month = feb,
	number = {2},
	pages = {L21},
	primaryclass = {astro-ph.HE},
	title = {{Saturation of the Filamentation Instability and Dispersion Measure of Fast Radio Bursts}},
	volume = {943},
	year = 2023}

@article{Gunn&Ostriker_71,
	adsurl = {https://ui.adsabs.harvard.edu/abs/1971ApJ...165..523G},
	author = {{Gunn}, James E. and {Ostriker}, Jeremiah P.},
	doi = {10.1086/150919},
	journal = {\apj},
	month = may,
	pages = {523},
	title = {{On the Motion and Radiation of Charged Particles in Strong Electromagnetic Waves. I. Motion in Plane and Spherical Waves}},
	volume = {165},
	year = 1971}

@article{Ghosh+22,
	adsurl = {https://ui.adsabs.harvard.edu/abs/2022ApJ...930..106G},
	archiveprefix = {arXiv},
	author = {{Ghosh}, Arka and {Kagan}, Daniel and {Keshet}, Uri and {Lyubarsky}, Yuri},
	doi = {10.3847/1538-4357/ac581d},
	eid = {106},
	eprint = {2111.00656},
	journal = {\apj},
	month = may,
	number = {2},
	pages = {106},
	primaryclass = {astro-ph.HE},
	title = {{Nonlinear Electromagnetic-wave Interactions in Pair Plasma. I. Nonrelativistic Regime}},
	volume = {930},
	year = 2022}

@article{Iwamoto+23,
	adsurl = {https://ui.adsabs.harvard.edu/abs/2023MNRAS.522.2133I},
	archiveprefix = {arXiv},
	author = {{Iwamoto}, Masanori and {Sobacchi}, Emanuele and {Sironi}, Lorenzo},
	doi = {10.1093/mnras/stad1100},
	eprint = {2304.03577},
	journal = {\mnras},
	month = jun,
	number = {2},
	pages = {2133-2144},
	primaryclass = {astro-ph.HE},
	title = {{Kinetic simulations of the filamentation instability in pair plasmas}},
	volume = {522},
	year = 2023}

@article{Beloborodov_20a,
	adsurl = {https://ui.adsabs.harvard.edu/abs/2020ApJ...896..142B},
	archiveprefix = {arXiv},
	author = {{Beloborodov}, Andrei M.},
	doi = {10.3847/1538-4357/ab83eb},
	eid = {142},
	eprint = {1908.07743},
	journal = {\apj},
	month = jun,
	number = {2},
	pages = {142},
	primaryclass = {astro-ph.HE},
	title = {{Blast Waves from Magnetar Flares and Fast Radio Bursts}},
	volume = {896},
	year = 2020}

@article{Blandford&Znajek77,
	adsurl = {https://ui.adsabs.harvard.edu/abs/1977MNRAS.179..433B},
	author = {{Blandford}, R.~D. and {Znajek}, R.~L.},
	doi = {10.1093/mnras/179.3.433},
	journal = {\mnras},
	month = may,
	pages = {433-456},
	title = {{Electromagnetic extraction of energy from Kerr black holes.}},
	volume = {179},
	year = 1977}

@article{Cerutti_20,
	adsurl = {https://ui.adsabs.harvard.edu/abs/2020arXiv200811462C},
	archiveprefix = {arXiv},
	author = {{Cerutti}, Beno{\^\i}t and {Philippov}, Alexander and {Dubus}, Guillaume},
	eid = {arXiv:2008.11462},
	eprint = {2008.11462},
	journal = {arXiv e-prints},
	month = aug,
	pages = {arXiv:2008.11462},
	primaryclass = {astro-ph.HE},
	title = {{Dissipation of the striped pulsar wind and non-thermal particle acceleration: 3D PIC simulations}},
	year = 2020}

@article{Cordes&Chatterjee19,
	adsurl = {https://ui.adsabs.harvard.edu/abs/2019ARA&A..57..417C},
	archiveprefix = {arXiv},
	author = {{Cordes}, James M. and {Chatterjee}, Shami},
	doi = {10.1146/annurev-astro-091918-104501},
	eprint = {1906.05878},
	journal = {\araa},
	month = aug,
	pages = {417-465},
	primaryclass = {astro-ph.HE},
	title = {{Fast Radio Bursts: An Extragalactic Enigma}},
	volume = {57},
	year = 2019}

@article{Lorimer+07,
	adsurl = {http://adsabs.harvard.edu/abs/2007Sci...318..777L},
	archiveprefix = {arXiv},
	author = {{Lorimer}, D.~R. and {Bailes}, M. and {McLaughlin}, M.~A. and {Narkevic}, D.~J. and {Crawford}, F.},
	doi = {10.1126/science.1147532},
	eprint = {0709.4301},
	journal = {Science},
	month = nov,
	pages = {777},
	title = {{A Bright Millisecond Radio Burst of Extragalactic Origin}},
	volume = 318,
	year = 2007}

@article{Lyubarsky08,
	adsurl = {http://adsabs.harvard.edu/abs/2008ApJ...682.1443L},
	archiveprefix = {arXiv},
	author = {{Lyubarsky}, Y.},
	doi = {10.1086/589435},
	eid = {1443-1449},
	eprint = {0804.2069},
	journal = {\apj},
	month = aug,
	pages = {1443-1449},
	title = {{Induced Scattering of Short Radio Pulses}},
	volume = 682,
	year = 2008}

@article{Metzger+19,
	adsurl = {https://ui.adsabs.harvard.edu/abs/2019MNRAS.485.4091M},
	archiveprefix = {arXiv},
	author = {{Metzger}, Brian D. and {Margalit}, Ben and {Sironi}, Lorenzo},
	doi = {10.1093/mnras/stz700},
	eprint = {1902.01866},
	journal = {\mnras},
	month = may,
	number = {3},
	pages = {4091-4106},
	primaryclass = {astro-ph.HE},
	title = {{Fast radio bursts as synchrotron maser emission from decelerating relativistic blast waves}},
	volume = {485},
	year = 2019}

@article{Petroff+19,
	adsurl = {https://ui.adsabs.harvard.edu/abs/2019A&ARv..27....4P},
	archiveprefix = {arXiv},
	author = {{Petroff}, E. and {Hessels}, J.~W.~T. and {Lorimer}, D.~R.},
	doi = {10.1007/s00159-019-0116-6},
	eid = {4},
	eprint = {1904.07947},
	journal = {\aapr},
	month = may,
	number = {1},
	pages = {4},
	primaryclass = {astro-ph.HE},
	title = {{Fast radio bursts}},
	volume = {27},
	year = 2019}

@article{goldreich_julian_69,
	adsurl = {http://adsabs.harvard.edu/abs/1969ApJ...157..869G},
	author = {{Goldreich}, P. and {Julian}, W.~H.},
	doi = {10.1086/150119},
	journal = {\apj},
	month = aug,
	pages = {869-+},
	title = {{Pulsar Electrodynamics}},
	volume = 157,
	year = 1969}

@book{Kruer+19,
  title={The physics of laser plasma interactions},
  author={Kruer, William},
  year={2019},
  publisher={crc Press}
}

@InProceedings{Fonseca+02,
author="Fonseca, R. A.
and Silva, L. O.
and Tsung, F. S.
and Decyk, V. K.
and Lu, W.
and Ren, C.
and Mori, W. B.
and Deng, S.
and Lee, S.
and Katsouleas, T.
and Adam, J. C.",
editor="Sloot, Peter M. A.
and Hoekstra, Alfons G.
and Tan, C. J. Kenneth
and Dongarra, Jack J.",
title="OSIRIS: A Three-Dimensional, Fully Relativistic Particle in Cell Code for Modeling Plasma Based Accelerators",
booktitle="Computational Science --- ICCS 2002",
year="2002",
publisher="Springer Berlin Heidelberg",
address="Berlin, Heidelberg",
pages="342--351",
isbn="978-3-540-47789-1"
}

@ARTICLE{Li+21b,
       author = {{Li}, Fei and {Decyk}, Viktor K. and {Miller}, Kyle G. and {Tableman}, Adam and {Tsung}, Frank S. and {Vranic}, Marija and {Fonseca}, Ricardo A. and {Mori}, Warren B.},
        title = "{Accurately simulating nine-dimensional phase space of relativistic particles in strong fields}",
      journal = {Journal of Computational Physics},
         year = 2021,
        month = aug,
       volume = {438},
          eid = {110367},
        pages = {110367},
          doi = {10.1016/j.jcp.2021.110367},
archivePrefix = {arXiv},
       eprint = {2007.07556},
 primaryClass = {physics.comp-ph},
       adsurl = {https://ui.adsabs.harvard.edu/abs/2021JCoPh.43810367L}
}

@ARTICLE{Nishiura+26,
       author = {{Nishiura}, Rei and {Kamijima}, Shoma F. and {Ioka}, Kunihito},
        title = "{Induced Scattering of Fast Radio Bursts in Magnetar Magnetospheres}",
      journal = {arXiv e-prints},
         year = 2026,
        month = jan,
          eid = {arXiv:2601.18865},
        pages = {arXiv:2601.18865},
          doi = {10.48550/arXiv.2601.18865},
archivePrefix = {arXiv},
       eprint = {2601.18865},
 primaryClass = {astro-ph.HE},
       adsurl = {https://ui.adsabs.harvard.edu/abs/2026arXiv260118865N}
}

@ARTICLE{Nishiura+25b,
       author = {{Nishiura}, Rei and {Kamijima}, Shoma F. and {Ioka}, Kunihito},
        title = "{Unified kinetic theory of induced scattering: Compton, Brillouin, and Raman processes in magnetized electron and positron pair plasma}",
      journal = {arXiv e-prints},
         year = 2025,
        month = oct,
          eid = {arXiv:2510.12869},
        pages = {arXiv:2510.12869},
          doi = {10.48550/arXiv.2510.12869},
archivePrefix = {arXiv},
       eprint = {2510.12869},
 primaryClass = {astro-ph.HE},
       adsurl = {https://ui.adsabs.harvard.edu/abs/2025arXiv251012869N}
}

@ARTICLE{Nishiura+25a,
       author = {{Nishiura}, Rei and {Kamijima}, Shoma F. and {Iwamoto}, Masanori and {Ioka}, Kunihito},
        title = "{Induced Compton scattering in magnetized electron and positron pair plasma}",
      journal = {\prd},
         year = 2025,
        month = mar,
       volume = {111},
       number = {6},
          eid = {063055},
        pages = {063055},
          doi = {10.1103/PhysRevD.111.063055},
archivePrefix = {arXiv},
       eprint = {2411.00936},
 primaryClass = {astro-ph.HE},
       adsurl = {https://ui.adsabs.harvard.edu/abs/2025PhRvD.111f3055N}
}

@ARTICLE{Kamijima+26,
       author = {{Kamijima}, Shoma F. and {Nishiura}, Rei and {Iwamoto}, Masanori and {Ioka}, Kunihito},
        title = "{One-dimensional PIC Simulation of Induced Compton Scattering in Magnetized Electron-Positron Pair Plasma}",
      journal = {arXiv e-prints},
         year = 2026,
        month = jan,
          eid = {arXiv:2601.01169},
        pages = {arXiv:2601.01169},
          doi = {10.48550/arXiv.2601.01169},
archivePrefix = {arXiv},
       eprint = {2601.01169},
 primaryClass = {astro-ph.HE},
       adsurl = {https://ui.adsabs.harvard.edu/abs/2026arXiv260101169K}
}

@ARTICLE{Li+21a,
       author = {{Li}, Fei and {Miller}, Kyle G. and {Xu}, Xinlu and {Tsung}, Frank S. and {Decyk}, Viktor K. and {An}, Weiming and {Fonseca}, Ricardo A. and {Mori}, Warren B.},
        title = "{A new field solver for modeling of relativistic particle-laser interactions using the particle-in-cell algorithm}",
      journal = {Computer Physics Communications},
         year = 2021,
        month = jan,
       volume = {258},
          eid = {107580},
        pages = {107580},
          doi = {10.1016/j.cpc.2020.107580},
archivePrefix = {arXiv},
       eprint = {2004.03754},
 primaryClass = {physics.comp-ph},
       adsurl = {https://ui.adsabs.harvard.edu/abs/2021CoPhC.25807580L}
}

@ARTICLE{Vay_2002,
       author = {{Vay}, Jean-Luc},
        title = "{Asymmetric Perfectly Matched Layer for the Absorption of Waves}",
      journal = {Journal of Computational Physics},
         year = 2002,
        month = dec,
       volume = {183},
       number = {2},
        pages = {367-399},
          doi = {10.1006/jcph.2002.7175},
       adsurl = {https://ui.adsabs.harvard.edu/abs/2002JCoPh.183..367V}
}

@ARTICLE{Esarey&Schroeder_09,
       author = {{Esarey}, E. and {Schroeder}, C.~B. and {Leemans}, W.~P.},
        title = "{Physics of laser-driven plasma-based electron accelerators}",
      journal = {Reviews of Modern Physics},
         year = 2009,
        month = jul,
       volume = {81},
       number = {3},
        pages = {1229-1285},
          doi = {10.1103/RevModPhys.81.1229},
       adsurl = {https://ui.adsabs.harvard.edu/abs/2009RvMP...81.1229E}
}

@ARTICLE{Tangtartharakul+25,
       author = {{Tangtartharakul}, Kavin and {Arefiev}, Alexey and {Lyutikov}, Maxim},
        title = "{Complete reflection of nonlinear electromagnetic waves in underdense pair plasmas enabled by dynamically formed Bragg-like structures}",
      journal = {arXiv e-prints},
         year = 2025,
        month = sep,
          eid = {arXiv:2509.06230},
        pages = {arXiv:2509.06230},
          doi = {10.48550/arXiv.2509.06230},
archivePrefix = {arXiv},
       eprint = {2509.06230},
 primaryClass = {physics.plasm-ph},
       adsurl = {https://ui.adsabs.harvard.edu/abs/2025arXiv250906230T}
}

@article{Clemmow_1974, 
        author={Clemmow, P. C.},
        title={Nonlinear waves in a cold plasma by Lorentz transformation}, 
        journal={Journal of Plasma Physics}, 
        year={1974}, 
        volume={12}, 
        DOI={10.1017/S0022377800025125}, 
        number={2}, 
        pages={297–317}
}

@ARTICLE{Schlegel+09,
       author = {{Schlegel}, T. and {Naumova}, N. and {Tikhonchuk}, V.~T. and {Labaune}, C. and {Sokolov}, I.~V. and {Mourou}, G.},
        title = "{Relativistic laser piston model: Ponderomotive ion acceleration in dense plasmas using ultraintense laser pulses}",
      journal = {Physics of Plasmas},
         year = 2009,
        month = aug,
       volume = {16},
       number = {8},
          eid = {083103},
        pages = {083103},
          doi = {10.1063/1.3196845},
       adsurl = {https://ui.adsabs.harvard.edu/abs/2009PhPl...16h3103S}
}

@ARTICLE{Robinson+09,
       author = {{Robinson}, A.~P.~L. and {Gibbon}, P. and {Zepf}, M. and {Kar}, S. and {Evans}, R.~G. and {Bellei}, C.},
        title = "{Relativistically correct hole-boring and ion acceleration by circularly polarized laser pulses}",
      journal = {Plasma Physics and Controlled Fusion},
         year = 2009,
        month = feb,
       volume = {51},
       number = {2},
          eid = {024004},
        pages = {024004},
          doi = {10.1088/0741-3335/51/2/024004},
       adsurl = {https://ui.adsabs.harvard.edu/abs/2009PPCF...51b4004R}
}

@article{Edwards+16,
  title = {Strongly Enhanced Stimulated Brillouin Backscattering in an Electron-Positron Plasma},
  author = {Edwards, Matthew R. and Fisch, Nathaniel J. and Mikhailova, Julia M.},
  journal = {Phys. Rev. Lett.},
  volume = {116},
  issue = {1},
  pages = {015004},
  numpages = {5},
  year = {2016},
  month = {Jan},
  publisher = {American Physical Society},
  doi = {10.1103/PhysRevLett.116.015004},
  url = {https://link.aps.org/doi/10.1103/PhysRevLett.116.015004}
}

@ARTICLE{Gruzinov2019,
       author = {{Gruzinov}, Andrei},
        title = "{Nonlinear scattering of Fast Radio Bursts}",
      journal = {arXiv e-prints},
         year = 2019,
        month = dec,
          eid = {arXiv:1912.08150},
        pages = {arXiv:1912.08150},
          doi = {10.48550/arXiv.1912.08150},
archivePrefix = {arXiv},
       eprint = {1912.08150},
 primaryClass = {astro-ph.HE},
       adsurl = {https://ui.adsabs.harvard.edu/abs/2019arXiv191208150G}
}

@ARTICLE{LyutikovGurarie2025,
       author = {{Lyutikov}, Maxim and {Gurarie}, Victor},
        title = "{Anderson self-localization of light in pair plasmas}",
      journal = {arXiv e-prints},
         year = 2025,
        month = sep,
          eid = {arXiv:2509.20594},
        pages = {arXiv:2509.20594},
          doi = {10.48550/arXiv.2509.20594},
archivePrefix = {arXiv},
       eprint = {2509.20594},
 primaryClass = {physics.plasm-ph},
       adsurl = {https://ui.adsabs.harvard.edu/abs/2025arXiv250920594L}
}

@ARTICLE{Macchi+2013,
       author = {{Macchi}, Andrea and {Borghesi}, Marco and {Passoni}, Matteo},
        title = "{Ion acceleration by superintense laser-plasma interaction}",
      journal = {Reviews of Modern Physics},
         year = 2013,
        month = apr,
       volume = {85},
       number = {2},
        pages = {751-793},
          doi = {10.1103/RevModPhys.85.751},
archivePrefix = {arXiv},
       eprint = {1302.1775},
 primaryClass = {physics.plasm-ph},
       adsurl = {https://ui.adsabs.harvard.edu/abs/2013RvMP...85..751M}
}

@BOOK{LandauLifshitz1975,
       author = {{Landau}, Lev Davidovich and {Lifshitz}, E.~M.},
        title = "{The classical theory of fields}",
         year = "1975",
         publisher = {Oxford Pergamon Press},
       adsurl = {https://ui.adsabs.harvard.edu/abs/1975ctf..book.....L}
}

@ARTICLE{Arrowsmith+24,
       author = {{Arrowsmith}, C.~D. and {Simon}, P. and {Bilbao}, P.~J. and {Bott}, A.~F.~A. and {Burger}, S. and {Chen}, H. and {Cruz}, F.~D. and {Davenne}, T. and {Efthymiopoulos}, I. and {Froula}, D.~H. and {Goillot}, A. and {Gudmundsson}, J.~T. and {Haberberger}, D. and {Halliday}, J.~W.~D. and {Hodge}, T. and {Huffman}, B.~T. and {Iaquinta}, S. and {Miniati}, F. and {Reville}, B. and {Sarkar}, S. and {Schekochihin}, A.~A. and {Silva}, L.~O. and {Simpson}, R. and {Stergiou}, V. and {Trines}, R.~M.~G.~M. and {Vieu}, T. and {Charitonidis}, N. and {Bingham}, R. and {Gregori}, G.},
        title = "{Laboratory realization of relativistic pair-plasma beams}",
      journal = {Nature Communications},
         year = 2024,
        month = jun,
       volume = {15},
          eid = {5029},
        pages = {5029},
          doi = {10.1038/s41467-024-49346-2},
archivePrefix = {arXiv},
       eprint = {2312.05244},
 primaryClass = {physics.plasm-ph},
       adsurl = {https://ui.adsabs.harvard.edu/abs/2024NatCo..15.5029A}
}

@ARTICLE{Chen&Fiuza_23,
       author = {{Chen}, Hui and {Fiuza}, Frederico},
        title = "{Perspectives on relativistic electron-positron pair plasma experiments of astrophysical relevance using high-power lasers}",
      journal = {Physics of Plasmas},
         year = 2023,
        month = feb,
       volume = {30},
       number = {2},
          eid = {020601},
        pages = {020601},
          doi = {10.1063/5.0134819},
       adsurl = {https://ui.adsabs.harvard.edu/abs/2023PhPl...30b0601C}
}

@ARTICLE{Palmer+11,
       author = {{Palmer}, Charlotte A.~J. and {Dover}, N.~P. and {Pogorelsky}, I. and {Babzien}, M. and {Dudnikova}, G.~I. and {Ispiriyan}, M. and {Polyanskiy}, M.~N. and {Schreiber}, J. and {Shkolnikov}, P. and {Yakimenko}, V. and {Najmudin}, Z.},
        title = "{Monoenergetic Proton Beams Accelerated by a Radiation Pressure Driven Shock}",
      journal = {\prl},
         year = 2011,
        month = jan,
       volume = {106},
       number = {1},
          eid = {014801},
        pages = {014801},
          doi = {10.1103/PhysRevLett.106.014801},
archivePrefix = {arXiv},
       eprint = {1006.3163},
 primaryClass = {physics.plasm-ph},
       adsurl = {https://ui.adsabs.harvard.edu/abs/2011PhRvL.106a4801P}
}

@ARTICLE{Haberberger+12,
       author = {{Haberberger}, Dan and {Tochitsky}, Sergei and {Fiuza}, Frederico and {Gong}, Chao and {Fonseca}, Ricardo A. and {Silva}, Luis O. and {Mori}, Warren B. and {Joshi}, Chan},
        title = "{Collisionless shocks in laser-produced plasma generate monoenergetic high-energy proton beams}",
      journal = {Nature Physics},
         year = 2012,
        month = jan,
       volume = {8},
       number = {1},
        pages = {95-99},
          doi = {10.1038/nphys2130},
       adsurl = {https://ui.adsabs.harvard.edu/abs/2012NatPh...8...95H}
}

@ARTICLE{Iwamoto&Ioka_26,
       author = {{Iwamoto}, Masanori and {Ioka}, Kunihito},
        title = "{Induced Scattering of Strong Waves in Pair Plasmas}",
      journal = {arXiv e-prints},
         year = 2026,
        month = apr,
          eid = {arXiv:2604.15798},
        pages = {arXiv:2604.15798},
archivePrefix = {arXiv},
       eprint = {2604.15798},
 primaryClass = {astro-ph.HE},
       adsurl = {https://ui.adsabs.harvard.edu/abs/2026arXiv260415798I}
}

\cleardoublepage

\section{End Matter}
\textit{Numerical simulation details---}Our numerical simulations use a highly accurate particle pusher \citep{Li+21b}, which is based on a class of exact solutions to the equations of motion, expressed as a function of proper time.
The shape of the macroparticles is represented with quadratic splines.  The EM field solver employs a second-order finite-difference method using a modified 16-coefficient stencil along the wave-propagation direction ($\hat{x}$). This implementation mitigates dispersion errors and spurious forces arising from the time staggering of electric and magnetic fields in the particle pusher \cite{Li+21a}. This choice requires the computational domain to have a finite number of cells in the transverse direction; specifically, the quadratic particle splines require at least 5 cells along the $\hat{y}$ direction. Still, since the 5-cell-wide domain width is much smaller than all other length scales relevant for this problem, our setup is effectively one-dimensional. We also apply a correction to the electric current in the Fourier space, so that the charge continuity equation remains
satisfied when using the modified stencil \cite{Li+21a} even when $a_0\gg1$. 

At $t=0$ the EM pulse is initialized in a vacuum region, located to the left of the space filled with plasma. The amplitude of this pulse is gradually increased and decreased over a length of $\ell_{\rm rise} = \ell_{\rm fall} = 10$ wavelengths using a Gaussian-like, 5th-order polynomial profile function of the form $f(x) = 10l(x)^3 - 15l(x)^4 + 6l(x)^5$, where $l(x) = (x - x_{\rm head})/\ell_{\rm rise}$ for the rising phase of the pulse (right) and $l(x) =(x_{\rm head} + \ell_{\rm flat} - x)/\ell_{\rm fall}$ for the falling phase (left). Here, $x_{\rm head}$ is the position of the right edge of the pulse, and $\ell_{\rm flat}=10^2-10^4$ wavelengths is the length of the `flat' part of the pulse (with strength parameter $a_0$), which is chosen to be sufficiently long to capture nonlinear propagation effects. We resolve the wavelength of the EM pulse with $\ge 100$ cells, the electron skin depth $c/\omp$ with $160-16000$ cells, and the Debye length $\lambda_{\rm D}=\sqrt{\theta_0}(c/\omp)$ with at least one cell for our fiducial case of $\theta_0=10^{-4}$. Our simulations employ 50 particles per cell (equally divided between electrons and positrons).
In the transition zone between the vacuum and the plasma, the density profile smoothly increases from 0 to 1 using a hyperbolic tangent function of the form $g(x) = 0.5\left[1 + \tanh{(x-1.1 x_{\rm head})}\right]$.
The choice of a smooth profile in the amplitude of the EM pulse and in the plasma density results in a gradual increase of the current and minimizes potential numerical artifacts during the early stages of interaction.

\begin{figure}[H]
\includegraphics[width=0.49\textwidth]{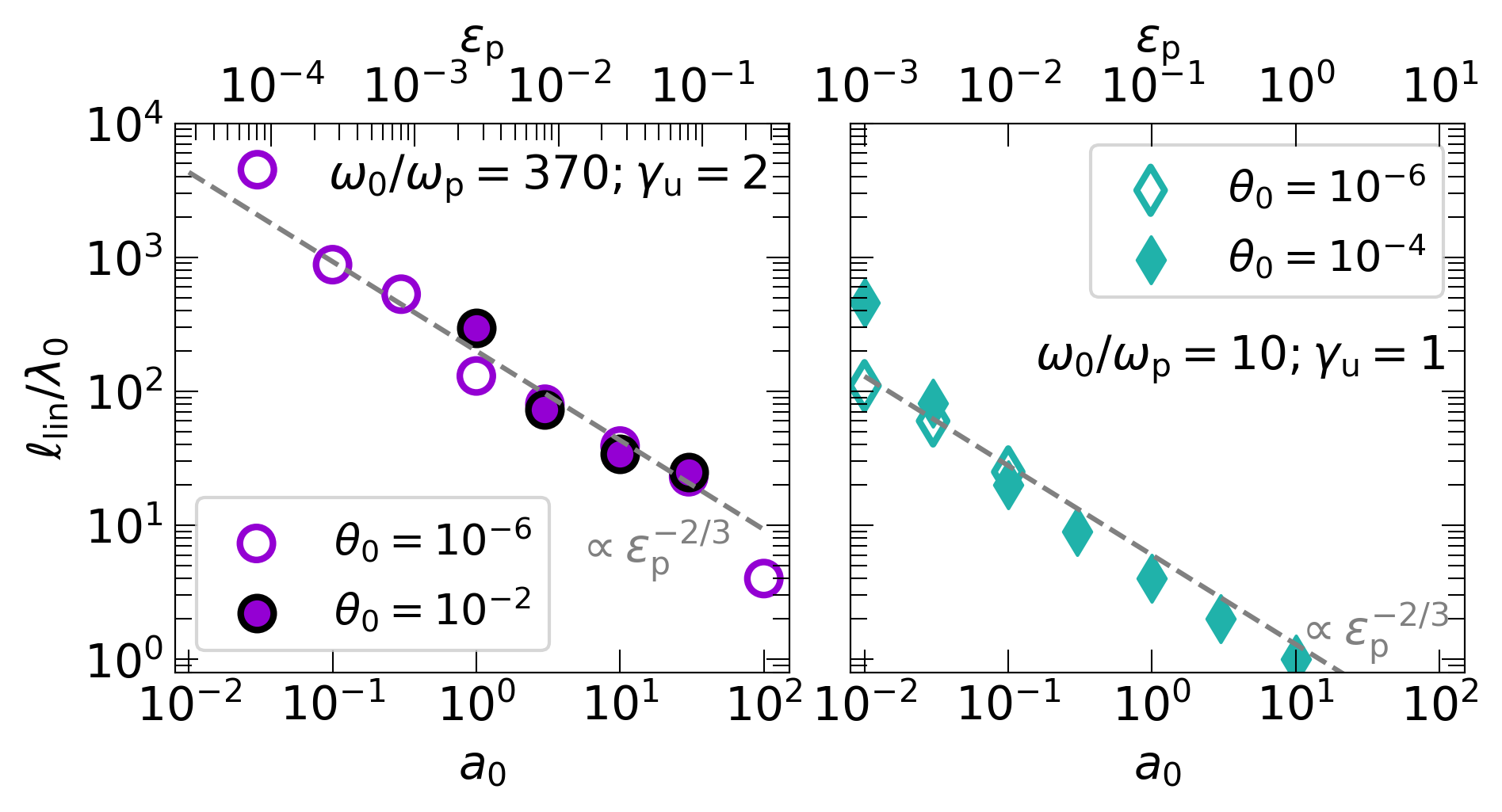}
\caption{\label{fig:lin_prop_length_gamma1temp}
Linear propagation length in units of the vacuum wavelength as a function of $a_0$ (bottom horizontal axis) and $\varepsilon_{\rm p}$ (top horizontal axis). The dashed lines show the $\varepsilon_{\rm p}^{-2/3}$ scaling. Left panel: simulations with the same $\gu=2$ and $\omo/\omp=370$, but different temperatures $\theta_0=10^{-6}$ (hollow markers) and $\theta_0=10^{-2}$ (black-bordered solid-filled markers). Right panel: simulations with the same $\gu=1$ and $\omo/\omp=10$, but different temperatures $\theta_0=10^{-6}$ (hollow markers) and $\theta_0=10^{-4}$ (solid-filled markers).
}
\end{figure}

\textit{Dependence of $\llinear$ on $\gu$ and $\theta_0$---}In Fig.~\ref{fig:lin_prop_length_gamma1temp}, we show the linear propagation length in units of the vacuum wavelength as a function of $a_0$ (bottom horizontal axis) and $\varepsilon_{\rm p}$ (top horizontal axis). The simulations shown in the left panel have the same $\gu=2$ and $\omo/\omp=370$, but different temperatures $\theta_0=10^{-6}$ (hollow markers) and $\theta_0=10^{-2}$ (black-bordered solid-filled markers). The simulations shown in the right panel have the same $\gu=1$ and $\omo/\omp=10$, but different temperatures $\theta_0=10^{-6}$ (hollow markers) and $\theta_0=10^{-4}$ (solid-filled markers). All results are consistent with the scaling $\llinear/\lambda_0\simeq\varepsilon_{\rm p}^{-2/3}$ predicted by Eq.~\eqref{eq:mainresult}.

\end{document}